\documentclass{appolb}
\usepackage{epsfig}
\title{NUCLEON SPIN STRUCTURE FUNCTIONS}

\author{Dorota Kotlorz
\address{Department of Physics, Technical University of Opole, Ozimska 75,
45-370 Opole, Poland, e-mail: {\tt dstrozik@po.opole.pl}}}

\begin{document}
\pagestyle{plain}
\eqsec
\maketitle

\begin{abstract}
The review of the nucleon spin structure functions problems
is presented. The perturbative QCD predictions for the small
$x$ behaviour of the nucleon spin structure functions is discussed.
The role of the resummation of the $\ln^2 1/x$ terms is
emphasised. Predictions for the nonsinglet structure function $g_1$
in case of a flat as well as a dynamical input are given. The
so called 'spin crisis' in the context of both theoretical and
experimental future hopes are also briefly discussed.
\end{abstract}


\section{Introduction}

Since 1988, when the famous EMC experiment \cite{b1} provided
surprising results, the polarised deep inelastic lepton-nucleon
scattering (DIS) became very interesting from experimental as well as
theoretical point of view. This experiment, in which longitudinally
polarised muons scattered on longitudinally polarised protons, brought
the conclusion that quarks are carrying only a small part of the
proton spin projection in the polarised proton. This result called
as 'spin crisis' is still a challenge for theoretical and experimental
research. The main questions should be answered are: how the nucleon
spin is distributed among its constituents: quarks and gluons and how
the dynamics of these constituent interactions depend on spin.
Solutions of these problems may be found within perturbative QCD
because they involve hard and semihard (short-distance) processes.
From experimental point of view there are many projects, which provide
or will provide a mechanism to probe the spin properties of nucleon
and should be a crucial test of QCD. The most important of these
projects are experiments in SLAC, DESY (HERMES, HERA), CERN (SMC)
(with polarised proton, deuteron and ${\rm He}^3$ targets),
CERN (COMPASS) (with polarised muon beams and longitudinally
polarised hydrogen ${\rm (NH}_3)$ and deuteron ${\rm (Li^6D)}$
targets. The main goal of all these experiments is to measure
the nucleon spin structure functions $g_1(x,Q^2)$ and $g_2(x,Q^2)$.
It can be done by measurement of the cross section asymmetry, where
two considered cases correspond to antiparallel or parallel spin
orientation of the longitudinally polarised lepton $(\mu,e)$ and
nucleon $(p,n,d)$ respectively. Recently the experiment data has
allowed to investigate the nucleon spin structure in the large range
of the kinematical variables: Bjorken $x$ and $Q^2$. The most
interesting, both theoretically and phenomenologically, is the region
of small $x$. Theoretical understanding of the small $x$
($x\sim 10^{-3}$ and less) behaviour of the polarised nucleon
structure function enables the correct estimation of $\Gamma_1$
momenta in sum rules. It is very important because present
experimental data do not cover however the whole very small $x$
region and the only way (at present) to know the nucleon spin
structure completely is extrapolation of large and medium $x$
results into the small $x$ region through the theoretical QCD
analysis. From the other side, future polarised experiments
in HERA \cite{b2} will enable spin DIS investigations in the very
small $x$ region: $x\sim 10^{-4}$ and less. Then theoretical
predictions would be verified by the experiment. These future spin
experiments would be a crucial test of theoretical analysis.
Description of the nucleon spin structure function $g_1$ within
perturbative QCD for small $x$ can be done in different frames
(in LO, NLO, $\ln 1/x$, $\ln^2 1/x$ etc. approximations) giving
different results for $g_1$ in this region. Thus the future comparison
of theoretical and experimental results could be definitive.
In the next section we shall briefly remind the sum rules and the
'spin crisis' problem. In section 3 we shall discuss the polarised
structure functions of nucleon in the small Bjorken $x$ region. We
shall emphasise the $\ln^2 1/x$ resummation which is significant
in this region. In point 4 the nonsinglet $g_1^{NS}(x,Q^2)$
predictions are presented. We show LO and unified LO+$\ln^2 1/x$
resummation results in case of a flat (nondynamical) and a dynamical
input parametrisation as well. We compare our results with recent SMC
data. Finally in conclusions we shall briefly discuss future
experimental hopes and possible scenario of solving the spin crisis
problem.

\section{Sum rules and the 'spin crisis'}

There are four basic nucleon structure functions: $F_1$, $F_2$ for
a spin independent case and $g_1$, $g_2$ for a polarised one, which
characterise the pointlike interaction between virtual hard photon
(Compton scattering) of $Q^2\gg\Lambda^2$ and hadron
constituents - partons in the deep inelastic lepton - hadron
scattering. In the parton model $F_1$, $F_2$ and $g_1$ have a very
simple form and interpretation. Thus
\begin{equation}\label{r2.1}
F_2(x)=x \sum\limits_{i=u,d,s,..} e_i^2[q_{i+}(x)+q_{i-}(x)]
\end{equation}
where $e_i$ is a charge of the i-flavour quark, $q_{i+}(x)$
$(q_{i-}(x))$ is the density distribution function of the
i-quark with the spin parallel (antiparallel) to the parent nucleon.
In the Bjorken limit
\begin{equation}\label{r2.2}
F_2(x)=2xF_1(x)
\end{equation}

\begin{equation}\label{r2.3}
g_1(x)=\frac12 \sum\limits_{i=u,d,s,..} e_i^2 \Delta q_i(x)
\end{equation}

\begin{equation}\label{r2.4}
\Delta q_i(x)=q_{i+}(x)-q_{i-}(x)
\end{equation}
Function $g_1(x,Q^2)$ is connected with the helicity of the nucleon
({\it i.e.} spin projection on the momentum direction). Thus the integral
\begin{equation}\label{r2.5}
\langle\Delta q_i\rangle =\int\limits_0^1 \Delta q_i(x) dx
\end{equation}
is simply a part of the nucleon helicity, carried by a quark of
i-flavour (i=u,d,s,..). The second spin dependent structure function
$g_2(x,Q^2)$, related to the transverse spin polarisation of the
nucleon has no simple meaning in the parton model. The main goal in
the deep inelastic lepton - nucleon scattering experiments with
polarised both the lepton and the nucleon particles is to find the
spin dependent structure function of nucleon $g_1(x,Q^2)$. This
measurement of $g_1(x,Q^2)$ provides the knowledge how the spin of
the nucleon is distributed among the partons: valence quarks
$u_v$,$d_v$, sea quarks $q_{sea}$ and gluons $g$. The experimental
measurement of $g_1(x,Q^2)$ is based on the measurement of the cross
section asymmetry factor $A$ \cite{b13}:
\begin{equation}\label{r2.6}
A=\frac{\sigma(++)-\sigma(+-)}{\sigma(++)+\sigma(+-)}
\end{equation}
where $\sigma(++)$, $\sigma(+-)$ correspond to the cases when the
spins of longitudinally polarised lepton ($\mu$ or $e$) and nucleon
($p$ or $n$ or $d$) are parallel or antiparallel. Finally,
$g_1(x,Q^2)$ can be determined through the relation
\begin{equation}\label{r2.7}
g_1(x,Q^2)=\frac{F_2(x,Q^2)A_1(x,Q^2)}{2x(1+R)}
\end{equation}
where
\begin{equation}\label{r2.8}
A_1=\frac{\sigma_{1/2}-\sigma_{3/2}}
{\sigma_{1/2}+\sigma_{3/2}}
\end{equation}
and
\begin{equation}\label{r2.9}
R=\frac{\sigma_L}{\sigma_T}
\end{equation}
The cross section $\sigma_{3/2}$ and $\sigma_{1/2}$ correspond to the
helicities 3/2 and 1/2 of the absorbed virtual photon. $\sigma_L$
and $\sigma_T$ are the absorption cross sections of longitudinal and
transverse virtual photons in the polarised scattering. Through
polarised DIS experiments with different targets: proton, neutron or
deuteron ones, it is possible to combine the $g_1$ results for these
different cases and hence to find out the spin dependent distribution
functions of partons ($\Delta u_v$, $\Delta d_v$, $\Delta q_{sea}$,
$\Delta g$) in the nucleon. The measurement of $g_1$ function enables
also verification of the sum rules, which play a very important role
as a test of QCD. The most significant of them are Bjorken and
Ellis-Jaffe sum rules \cite{b7}. Both sum rules are related with the
nucleon spin structure functions and their estimation requires the
knowledge of first moments of $g_1$ for proton, neutron and deuteron
$\Gamma_1^p(Q^2)$, $\Gamma_1^n(Q^2)$, $\Gamma_1^d(Q^2)$:
\begin{equation}\label{r2.10}
\Gamma_1^i(Q^2)=\int\limits_0^1 g_1^i(x,Q^2) dx
\end{equation}
To test experimentally the Bjorken and Ellis-Jaffe sum rules it is
necessary to know function $g_1(x,Q^2)$ for certain $Q^2$ and in the
entire range of $x$: $x\in(0;1)$. This data collection of $g_1$ for
arbitrary values of $Q^2$ and $x$ variables is however impossible
because of technical constraint in the experiments. The broad ranges
of $x$ and $Q^2$ (where $Q^2=-q^2$, $q$ is the four momentum transfer
between lepton and  hadron and $W^2=Q^2(1/x-1)$ is total CM energy
squared) cannot be reached independently. The accessible at present
kinematical region in the polarised fixed target HERMES experiment
in HERA is $0.004<x<1$, $0.2<Q^2<20~{\rm GeV}^2$ \cite{b26}. It must
be however emphasised, that HERA will extend the present region by
two orders of magnitude both in $x$ and $Q^2$: $x$ down to
$6\cdot10^{-5}$ and $Q^2$ up to $2\cdot10^4~{\rm GeV}^2$. It will take
place in the future experiment with polarised electron and proton
beams \cite{b8}. Hence physicists hope for a new very interesting
field of experimental investigations. Broader range of $x$ and $Q^2$
in the future polarised DIS experiments in HERA will enable more
precise knowledge of the proton spin structure and the test of QCD
predictions for the polarised structure functions of the nucleon.
The accessible region of $x$ and $Q^2$ in future polarised experiments
in HERA is roughly presented in Fig.1 \cite{b8}.
\begin{figure}[ht]
\begin{center}
\includegraphics[]{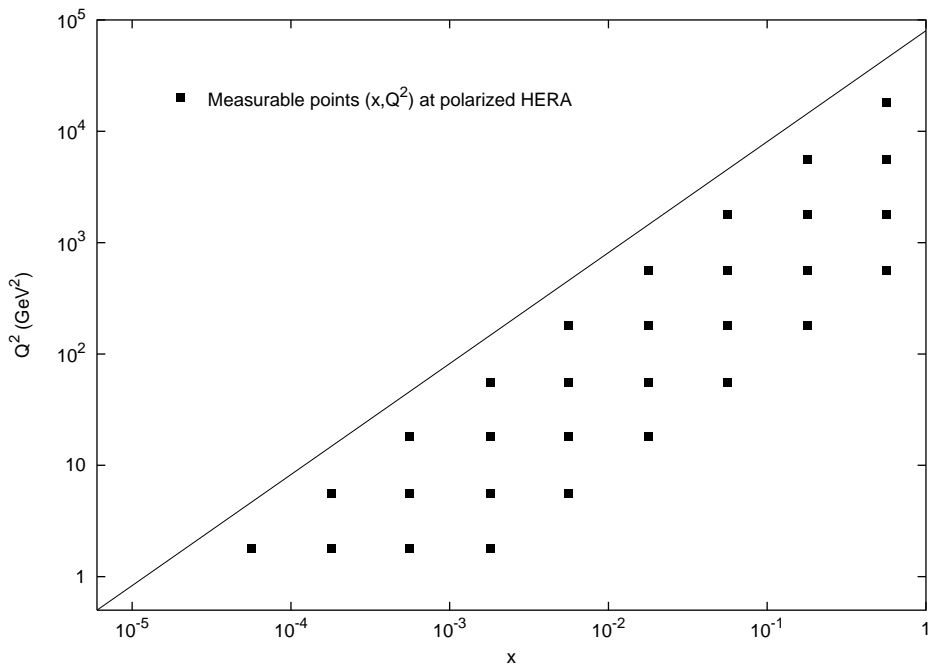}
\caption{Available ($x,Q^2$) region for polarised experiments in HERA
(below the sloped line).}
\end{center}
\end{figure}
Perturbative QCD provides methods for theoretical analysis of structure
functions for $Q^2\gg\Lambda^2$ ($\Lambda \sim 200$ MeV).
When the coupling constant of strong
interactions $\alpha_s(Q^2)$ becomes large ($\alpha_s\sim 1$)
perturbative calculations are not applicable.
QCD as a theory with asymptotic freedom
improves the naive parton model by permission
for a weak $Q^2$ dependence of structure
functions $F_1$, $F_2$, $g_1$, $g_2$. This scaling violation
is caused by strong interactions between
partons (quarks and gluons). The larger $Q^2$
({\it i.e.} smaller distance between partons) the
weaker interaction between partons.
Perturbative QCD enables $Q^2$ evolution of
structure functions, where the values of the
structure functions at some fixed scale $Q^2=Q_0^2$
are taken from the experiment and used as an input in evolution
equations. As it was already mentioned above, the most important test
of QCD in the spin dependent DIS phenomena is verification of Bjorken
(BSR) and Ellis-Jaffe (EJSR) sum rules. BSR is related to the
isovector axial current SU(3) flavour symmetry in the nucleon $\beta$
decay while EJSR results from the octet axial current SU(3) symmetry
in the $\beta$ decay of hyperons. In the parton model BSR and EJSR
have forms:
\begin{equation}\label{r2.11}
BSR:~~~a_3=g_A
\end{equation}
where
\begin{equation}\label{r2.12}
a_3=\int\limits_0^1(\Delta u+\Delta \bar{u}-\Delta d-\Delta \bar{d})dx
\end{equation}
and $g_A\approx 1.257$ is $\beta$ decay structure constant for neutron;
\begin{equation}\label{r2.13}
EJSR:~~~a_8=3F-D
\end{equation}
where
\begin{equation}\label{r2.14}
a_8=\int\limits_0^1 g_1^{octet}(x)dx=\int\limits_0^1(\Delta u+
\Delta \bar{u}+\Delta d+\Delta \bar{d}-2\Delta s-2\Delta \bar{s})dx
\end{equation}
$\Delta q (q=u,d,s,..)$ are polarised distribution functions of quarks
and antiquarks (\ref{r2.4}) and $F,D$ are axial coupling constants,
estimated from the weak $\beta$ decays of hyperons:
$3F-D\approx 0.579$. Taking into account the QCD corrections one can
obtain:
\begin{equation}\label{r2.15}
a_{0,3,8}\rightarrow a_{0,3,8}(1-corr(\alpha))
\end{equation}
where $corr(\alpha)$ is just a QCD correction, calculated in a
perturbative way \cite{b13}:
\begin{equation}\label{r2.16}
corr(\alpha)\approx (\alpha_s/\pi)+3.58(\alpha_s/\pi)^2+
20.22(\alpha_s/\pi)^3+130(\alpha_s/\pi)^4
\end{equation}
$\alpha_s(Q^2)$ is the running coupling constant of strong interaction
and in LO approximation has a form
\begin{equation}\label{r2.17}
\alpha_s(Q^2)=\frac{12\pi}{(33-2N_f)\ln\frac{Q^2}{\Lambda^2}}
\end{equation}
where $\Lambda=232$ MeV is a scale parameter of QCD.
$a_0$ in (\ref{r2.15}) denotes the singlet part of axial current:
\begin{equation}\label{r2.18}
a_0=\int\limits_0^1 g_1^{singlet}(x)dx=\int\limits_0^1\Delta\Sigma dx
\end{equation}
For three flavours of quarks ($N_f=3$) $\Delta\Sigma$ is given by
\begin{equation}\label{r2.19}
\Delta q_S\equiv\Delta\Sigma\equiv\Delta u+\Delta \bar{u}+\Delta d+
\Delta \bar{d}+\Delta s+\Delta \bar{s}
\end{equation}
Thus, taking into account QCD corrections (\ref{r2.15}) one can
rewrite the BSR (\ref{r2.11}) in the form
\begin{equation}\label{r2.20}
\Gamma_1^p-\Gamma_1^n\equiv\int\limits_0^1 (g_1^p-g_1^n)dx=
\frac16 g_A(1-corr(\alpha))
\end{equation}
Bjorken and Ellis-Jaffe sum rules imply the following expressions for first moments of spin
structure functions $g_1$ of the nucleon:
\begin{equation}\label{r2.21}
\Gamma_1^p\equiv\int\limits_0^1 g_1^p(x)dx=(\frac{a_3}{12}+\frac{a_8}{36}
+\frac{a_0}{9})(1-corr(\alpha))
-\frac{N_f}{18\pi}\alpha_s(Q^2)\langle\Delta g(Q^2)\rangle
\end{equation}
\begin{equation}\label{r2.22}
\Gamma_1^n\equiv\int\limits_0^1 g_1^n(x)dx=(-\frac{a_3}{12}+\frac{a_8}{36}
+\frac{a_0}{9})(1-corr(\alpha))
-\frac{N_f}{18\pi}\alpha_s(Q^2)\langle\Delta g(Q^2)\rangle
\end{equation}
\begin{eqnarray}\label{r2.23}
\Gamma_1^d\equiv (1-\frac32\omega_D)\int\limits_0^1 g_1^d(x)dx=
[(\frac{a_8}{36}+\frac{a_0}{9})(1-corr(\alpha))\nonumber\\
-\frac{N_f}{18\pi}\alpha_s(Q^2)\langle\Delta g(Q^2)\rangle]
(1-\frac32\omega_D)
\end{eqnarray}
where $\Delta g$ is the spin dependent distribution function of gluons
and $\langle\Delta g\rangle$ is defined similarly as for quarks
(\ref{r2.5}). Parameter $\omega_D\sim 0.058$ denotes probability that
deuteron is in D state. QCD corrections $\alpha_s^n$ in formulae
(\ref{r2.20})-(\ref{r2.23}) enable us to compare theoretical
perturbative QCD predictions for the structure function $g_1$ and its
first moment $\Gamma_1$ with the experimental results (\ref{r2.7}).
Thus one can easily read from (\ref{r2.20}), (\ref{r2.16}) and
(\ref{r2.17}) that BSR at $Q^2=10~{\rm GeV}^2$ for $\Lambda =200$ MeV
up to $\alpha_s^4$ correction gives
\begin{equation}\label{r2.24}
THEORY:~~~\Gamma_1^p-\Gamma_1^n=0.185
\end{equation}
This result is in a good agreement with the most recent SMC and SLAC
measurements \cite{b13},\cite{b14}, which yield for
$Q^2=10~{\rm GeV}^2$ the quantity
\begin{equation}\label{r2.25}
EXPERIMENT:~~~\Gamma_1^p-\Gamma_1^n=0.195\pm 0.029
\end{equation}
It means that the Bjorken sum rule BSR is fulfilled within the
experimental errors. For the Ellis-Jaffe sum rule EJSR however, there
is a significant disagreement between theory and experiment.
Theoretical value for EJSR from the simplest version of (\ref{r2.21}),
where the gluon contribution is neglected and the strange sea quarks
are unpolarised $(\Delta s=0)$ gives \cite{b13}
\begin{equation}\label{r2.26}
THEORY:~~~\Gamma_1^p(Q^2=10.7~{\rm GeV}^2)=0.171
\end{equation}
while the EMC result \cite{b1} is
\begin{eqnarray}\label{r2.27}
EXPERIMENT:~~~\Gamma_1^p(Q^2=10.7~{\rm GeV}^2)=\nonumber\\
=0.126\pm 0.010(statist)\pm 0.015(system)
\end{eqnarray}
Comparing (\ref{r2.26}) and (\ref{r2.27}) one can see, that the EJSR
prediction disagrees with experimental data by 2.6 standard
deviations. From the latest experimental analyses \cite{b11},
\cite{b12},\cite{b14} it can be read that BSR is always fulfilled
within the experimental errors while EJSR is broken at the level of
about 3 standard deviations. This fact is confirmed by many polarised
DIS experiments: SLAC in CERN with the polarised electron beam
scattered on polarised proton $(p)$, neutron $(n)$ or deuteron $(d)$
targets, SMC at CERN with muons $\mu^+$ and $p$,$d$ targets and HERMES
at DESY experiments with positons $e^+$ and $n$,$p$,$d$ targets. The
characteristic of kinematic variables for experimental groups EMC,
SLAC, SMC and HERMES is presented below in Tab.1 \cite{b12}.
\begin{table}[h]
\title{TABLE I}
\begin{center}
\begin{tabular}{|c|c|c|c|c|c|}
\hline
Experiment & Beam & Lepton & Smallest $x$ & Average
& Nucleon\\
 & & energy & $Q^2>1 $ & $Q^2$ & target\\
 & & E (GeV) & (${\rm GeV^2}$) & (${\rm GeV^2}$) & \\
\hline \hline
EMC & $\mu^+$ & 100-200 & 0.01 & 10.7 & p\\ \hline
SMC & $\mu^+$ & 100-190 & 0.003 & 10 & p,d\\ \hline
SLAC E-142 & $e^-$ & 19-25 & 0.03 & 2 & n\\ \hline
SLAC E-143 & $e^-$ & 10-29 & 0.029 & 3 & p,d\\ \hline
SLAC E-154 & $e^-$ & 48.3 & 0.014 & 5 & n\\ \hline
SLAC E-155 & $e^-$ & 48.3 & 0.014 & 5 & p,d\\ \hline
HERMES & $e^+$ & 27.5 & 0.023 & 2.3 & n,p,d\\ \hline
\end{tabular}
\caption{The kinematic variables in DIS for different experimental groups}
\end{center}
\end{table}
The experimental data of $g_1$, $\Gamma_1$ and
$\langle\Delta q_i\rangle$ (\ref{r2.5}) are widely reviewed {\it e.g.}
in \cite{b6},\cite{b16}. The main problem nowadays is to find out how
the spin of the proton is distributed among partons: valence quarks,
sea quarks and gluons. The part of the proton spin carried by the
parton $p$, where $p$ denotes $u_{val}$, $d_{val}$, $u_{sea}$,
$d_{sea}$, $s_{sea}$, antiquarks, $g$ is given by (\ref{r2.28}).
\begin{equation}\label{r2.28}
\langle\Delta p\rangle =\int\limits_0^1 \Delta p dx
\end{equation}
and
\begin{equation}\label{r2.29}
\Delta q=\Delta q_{val}+\Delta q_{sea}
\end{equation}
\begin{equation}\label{r2.30}
\Delta q_{sea}=\Delta\bar{q}_{sea}\equiv\Delta\bar{q}
\end{equation}
hence
\begin{equation}\label{r2.31}
\Delta q_{val}=\Delta q-\Delta\bar{q}
\end{equation}
The method of the extraction $\Delta p$ and then the important
quantity $\langle\Delta p\rangle$ from experimental data is as
follows:\\
1. Polarised distribution functions $\Delta p(x,Q^2)$ are parametrised
at the given low scale $Q_0^2$ ({\it e.g.} $Q_0^2=1~{\rm GeV}^2)$ in the
form:
\begin{equation}\label{r2.32}
\Delta p(x,Q_0^2)=N\eta_fx^{\alpha_f}(1-x)^{\beta_f}(1
+\gamma_fx^{\delta_f})
\end{equation}
where
\begin{equation}\label{r2.33}
\eta_f=\int\limits_0^1 \Delta p(x,Q_0^2)dx
\end{equation}
$N$ is a normalisation factor and $\alpha_f$, $\beta_f$, $\gamma_f$,
$\delta_f$ are parameters.\\
2. From QCD evolution equations ({\it see} Appendix A)
(\ref{rA.1})-(\ref{rA.3}) one can get in NLO approximation
$\Delta p(x,Q^2)$ functions for $(x,Q^2)$ set, available in
experiments.\\
3. Using the spin dependent distribution functions $\Delta p$, one can
calculate the spin structure function $g_1(x,Q^2)$ from the formula:
\begin{eqnarray}\label{r2.34}
g_1(x,Q^2)=\frac12\langle e^2\rangle\int\limits_x^1 \frac{dy}{y}[
C_S^q(x/y,\alpha_s(Q^2))\Delta\Sigma(y,Q^2)+\nonumber\\
2N_fC^g(x/y,\alpha_s(Q^2))\Delta g(y,Q^2)
+C_{NS}^q(x/y,\alpha_s(Q^2))\Delta q_{NS}(y,Q^2)]
\end{eqnarray}
where
\begin{equation}\label{r2.35}
\langle e^2\rangle=\frac{1}{N_f}\sum\limits_k^{N_f}e_k^2
\end{equation}
$\Delta q_S\equiv\Delta\Sigma$ (\ref{r2.19}) is singlet and
$\Delta q_{NS}$ is the nonsinglet polarised distribution function of
quarks:
\begin{equation}\label{r2.36}
\Delta q_{NS}\equiv\sum\limits_{i=1}^{N_f}(\frac{e_i^2}{\langle e^2\rangle}
-1)(\Delta q_i+\Delta\bar{q_i})
\end{equation}
4. Now one can compare the calculated $g_1$ (\ref{r2.34}) with
experimental data for $g_1$ (\ref{r2.7}) and fit free parameters
($\alpha_f$, $\beta_f$, $\gamma_f$, $\delta_f$) of input
parametrisations (\ref{r2.32}) in such a way to minimise the $\chi^2$
for all experimental points $(x,Q^2)$.\\
In this way it was found, as it was already mentioned above, that all
experiments confirm the validity of the Bjorken sum rule while the
Ellis-Jaffe sum rule is violated. After extraction of spin dependent
distribution functions $\Delta p$ from experimental data it was
possible to determine the value $\int\Delta\Sigma(x,Q^2))dx$, which is
a part of the proton spin carried by quarks. The result was surprising
because it turned out that quarks carry only a very small part of the
proton spin:
\begin{equation}\label{r2.37}
\int\limits_0^1 \Delta\Sigma (x,Q^2)dx\approx 0.17\pm 0.17
\end{equation}
and the strange quarks carry much larger part of the proton spin that
it follows from the simple parton model:
\begin{equation}\label{r2.38}
\int\limits_0^1 \Delta s(x,Q^2)dx\approx -0.14
\end{equation}
This result known since EMC experiments \cite{b1} as a "spin crisis"
problem should be correctly solved within QCD. Theoretical analysis
should provide a proper interpretation of the EMC results. The spin
problem is in fact not the "spin crisis" but the problem of
understanding how the nucleon spin is composed of parton spins?
Disagreement of experimental data with theoretical predictions,
emerging in EJSR violation, shows that the simple quark model is
definitely inadequate to the proper description of the nucleon spin
structure. The "spin crisis" has arisen because experimental results
did not confirm the naive quark model in which the proton spin is
carried only by valence quarks {\it i.e.}:
\begin{equation}\label{r2.39}
\int\limits_0^1 \Delta\Sigma dx=\int\limits_0^1(\Delta u_{val}
+\Delta d_{val})dx=1
\end{equation}
In this model it was assumed that strange sea quarks as well as gluons
do not or almost do not contribute to the nucleon spin. Using this
assumption, the input distribution functions (\ref{r2.32}) were
parametrised and then used in the fitting of experimental data. So to
remove the spin problem it is necessary to revise the previous
knowledge about the spin distribution in the nucleon. It would be
helpful with this aim to consider the following facts and assumptions
\cite{b6}:\\
1. The anomalous dimension of the first moment of gluons is equal to
-1:
\begin{equation}\label{r2.40}
\int\limits_0^1 \Delta g(x,Q^2)dx\sim \ln Q^2~~~for~Q^2\rightarrow\infty
\end{equation}
It means that
\begin{equation}\label{r2.41}
\lim_{Q^2 \to \infty}[\alpha_s(Q^2)\int\limits_0^1 \Delta g(x,Q^2)dx]=const
\end{equation}
and gluons give a nonvanishing correction of order $\alpha_s$ to EJSR
(\ref{r2.21})-(\ref{r2.23}), when $Q^2\rightarrow\infty$ (opposed to
the first moment of quarks, which the anomalous dimension is zero).
This gluon anomalous dimension via evolution equations
(\ref{rA.1})-(\ref{rA.3}) changes also the spin dependent distribution
functions of quarks.\\
2. Gluons can carry a very large part of the nucleon spin. If we
assume that
\begin{equation}\label{r2.42}
\int\limits_0^1 \Delta g(x,Q^2=10)dx\approx 2
\end{equation}
then \cite{b13}:
\begin{equation}\label{r2.43}
\int\limits_0^1 \Delta s(x,Q^2)dx=-0.06\pm 0.06
\end{equation}
\begin{equation}\label{r2.44}
\int\limits_0^1 \Delta\Sigma (x,Q^2)dx=0.42\pm 0.17
\end{equation}
3. The polarised sea quarks distribution function is large and
negative, so it cancels most of the valence quark contribution to the
nucleon spin.\\
4. The orbital momenta of the nucleon constituents $L_z(Q^2)$ is large
and it is a lacking part of the nucleon spin, according to the total
nucleon spin law conservation:
\begin{equation}\label{r2.45}
\frac12\langle\Delta\Sigma\rangle +\langle\Delta g\rangle +L_z(Q^2)
=\frac12
\end{equation}
The problem of the "spin crisis" is still open. Its solution requires
most of all the knowledge about polarised distribution functions of
gluons. Theoretical analyses QCD as well as future polarised
experiments at HERA \cite{b8}, COMPASS \cite{b15} and RHIC \cite{b21}
should better determine the gluon spin function behaviour,
particularly in the small Bjorken $x$ region, what is a crucial point
to overcome the "spin crisis".

\section{Spin structure functions in the small Bjorken $x$ region}

Determination of the nucleon spin structure functions in the small
Bjor\-ken $x$ region is very important from both theoretical and
experimental point of view. Because of technical limit, present
experiments do not give any information about small $x$ region
$(x\sim 10^{-4}, 10^{-5})$ and therefore there are still the
uncertainties in the determination of parton distribution functions
(in particular the gluons) in this region. Theoretical analyses, based
on the perturbative QCD, allow to calculate the nucleon structure
function within some approximations ($Q^2$LO, $Q^2$NLO, $\ln 1/x$
{\it etc.}). Choice of some particular approximation depends of course
on the region of its application and the basic criterion is agreement
of the theoretical predictions with experimental data. Thus the small
$x$ behaviour of the nucleon spin structure functions implied by QCD
can be via sum rules (BSR, EJSR) tested experimentally. Moreover, the
aim of the QCD analysis is to yield an adequate, compact description
of the nucleon structure functions in the whole range of $x$: for the
values of $x$, which are available in experiment and which are not as
well. The small $x$ region is also a challenge for QCD analysis,
because theoretical predictions of the structure function
$g_1^p(x,Q^2)$ at low $x$ are relevant for the future polarised HERA
measurements \cite{b8}.

Structure functions for small $x$ and fixed $Q^2$ are connected with
the virtual Compton scattering total cross-section at very high energy
$W^2\rightarrow\infty$:
\begin{equation}\label{r3.1}
W^2=Q^2(\frac1x -1)
\end{equation}
where $W$ is the total CM energy. The small value of $x$
($x\rightarrow 0$) corresponds by definition to the Regge limit and
therefore the small $x$ behaviour of structure functions can be
described using the Regge pole exchange model \cite{b2}. The Regge
theory predicts, that spin dependent structure functions $g_1^{p,n,d}$
in the small $x$ region behave as
\begin{equation}\label{r3.2}
g_1^{p,n,d}\sim x^{-\alpha}
\end{equation}
where $\alpha$ denotes the axial vector meson trajectory and lies in
limits:
\begin{equation}\label{r3.3}
-0.5\le\alpha\le 0
\end{equation}
The experimental data from HERA confirm such a Regge behaviour of
structure functions (\ref{r3.2}) but only in the low $Q^2$ region
$Q^2\le\Lambda^2$ ($\Lambda^2\approx 200$ MeV) {\it i.e.} in the
region, where the perturbative methods are not applicable. At larger
$Q^2$, because of parton interaction, the structure functions undergo
the $Q^2$ evolution and their behaviour, implied by perturbative QCD
is more singular than that, coming from the Regge picture. This fact
is also in agreement with experiments of unpolarised as well as
polarised DIS. It is well known at present, that for $x\rightarrow 0$
the Regge behaviour $x^{-\alpha}$ ($-0.5\le\alpha\le 0$) is less
singular than the perturbative QCD predictions for all of parton
distributions except unpolarised, nonsinglet (valence) quarks
$q_{NS}$.\\
There are two basic approaches of perturbative analysis of unpolarised
structure functions in the small $x$ region: a resummation of the GLAP
$Q^2$ logarithms \cite{b3},\cite{b10},\cite{b16}:
\begin{equation}\label{r3.4}
\sum\limits_{n,m} [\alpha_s(Q^2)]^n(\ln Q^2)^m
\end{equation}
or a resummation of $1/x$ logarithms BFKL \cite{b4}:
\begin{equation}\label{r3.5}
\sum\limits_{n,m} [\alpha_s(Q^2)]^n(\ln\frac1x )^m
\end{equation}
In the latest experimental data analyses of DIS the standard approach
is based on the QCD evolution equations within next-to leading (NLO)
approximation \cite{b3},\cite{b10}. This approach is appropriate for
polarised as well as unpolarised nucleon structure functions. It has
to be emphasised, that the changes, which  appear by the transition
from the LO approach to the NLO one are different in the case of
polarised and unpolarised structure functions. Splitting functions
$P_{ij}^{(0)}(x)$ and $\Delta P_{ij}^{(0)}(x)$, governing the
evolution of quark and gluon distribution functions in the LO
approximation have a form (\ref{rA.7})-(\ref{rA.10}) in the spin
independent case and a form (\ref{rA.11})-(\ref{rA.14}) in the spin
dependent case respectively. In the small $x$ limit $x\rightarrow 0$
they take a LO form:
\begin{equation}\label{r3.6}
P_{qq}^{(0)}(x)\sim\frac43;~~P_{qG}^{(0)}(x)\sim\frac12;~~
P_{Gq}^{(0)}(x)\sim\frac{8}{3x};~~P_{GG}^{(0)}(x)\sim\frac3x;
\end{equation}
\begin{equation}\label{r3.7}
\Delta P_{qq}^{(0)}(x)\sim\frac43;
~~\Delta P_{qG}^{(0)}(x)\sim -\frac12;
~~\Delta P_{Gq}^{(0)}(x)\sim\frac83;
~~\Delta P_{GG}^{(0)}(x)\sim 9
\end{equation}
In the NLO approximation, functions $P_{ij}^{(1)}(x)$ and
$\Delta P_{ij}^{(1)}(x)$ have the following small $x$ behaviour
\cite{b16},\cite{b17}:
\begin{equation}\label{r3.8}
P_{qq}^{(1)}(x)\sim\frac1x;~~P_{qG}^{(1)}(x)\sim\frac1x;~~
P_{Gq}^{(1)}(x)\sim -\frac1x;~~P_{GG}^{(1)}(x)\sim -\frac1x;
\end{equation}
\begin{equation}\label{r3.9}
\Delta P_{qq}^{(1)}(x),~\Delta P_{qG}^{(1)}(x),
~\Delta P_{Gq}^{(1)}(x),~\Delta P_{GG}^{(1)}(x)\sim a+b\ln x+c\ln^2x
\end{equation}
where $a$,$b$,$c$ are constants. From the comparison of
(\ref{r3.6})-(\ref{r3.9}) one can read an interesting information
about the small $x$ behaviour of polarised and unpolarised nucleon
structure functions. First: comparing (\ref{r3.6}) with (\ref{r3.8})
{\it i.e.} spin independent $P_{ij}(x)$ function in LO and NLO
approximations, one can notice, that the singular terms $1/x$ appear
in $P_{qq}^{(1)}(x)$ and $P_{qG}^{(1)}(x)$ while they do not in
$P_{qq}^{(0)}(x)$ and $P_{qG}^{(0)}(x)$ respectively. This means that
the evolution of unpolarised, fermion distribution functions at small
$x$ is completely dominated by the NLO terms in the perturbative
calculation. Secondly: the singular $1/x$ terms in
$P_{Gq}^{(1)}(x)$ and $P_{GG}^{(1)}(x)$ are opposite in sign with
respect to the corresponding LO terms:
 $P_{Gq}^{(0)}(x)$ and $P_{GG}^{(0)}(x)$. Thus, the rapid growth of
the unpolarised gluon distribution function at small $x$, well known
from LO analyses, will be dampened by the NLO contribution. Thirdly:
all spin dependent splitting functions $\Delta P_{ij}^{(1)}(x)$ are
nonsingular (without $1/x$ terms). (An exact form  of
$\Delta P_{ij}^{(1)}(x)$ is given {\it e.g.} in \cite{b16}.) It means,
in general, that in the small $x$ region spin dependent parton
distribution functions are by a factor of $x$ less singular than the
corresponding unpolarised distributions. Knowledge of the nucleon
structure function at small $x$ is extremely important: $g_1(x,Q^2)$
results for this region enter into integrals for moments
$\Gamma_1^{p,n,d}$ (\ref{r2.10}) and hence into the Bjorken and
Ellis-Jaffe sum rules (\ref{r2.20})-(\ref{r2.23}). From the other side,
lack of the very small $x$ ($x\le 10^{-4}$) experimental data causes
that the all knowledge of structure functions in limit small $x$
($x\rightarrow 0$) comes almost entirely from theoretical QCD
analyses. Therefore perturbative (and nonperturbative as well) QCD
predictions in the small $x$ region are at present of great
importance.\\
Evolution equations GLAP for unpolarised gluon distribution functions
$g(x,Q^2)$ and unpolarised singlet quark distributions
$\Sigma (x,Q^2)$
\begin{equation}\label{r3.10}
\Sigma (x,Q^2)=\sum\limits_{i=1}^{N_f}[q_i(x,Q^2)+\bar{q_i}(x,Q^2)]
\end{equation}
lead to the following behaviour \cite{b18}:
\begin{equation}\label{r3.11}
xg(x,Q^2)\sim \sigma^{-1/2}e^{2\gamma\sigma -\delta\zeta}
(1+\sum\limits_{i=1}^n\varepsilon^i\rho^{i+1}\alpha_s^i)
\end{equation}
\begin{equation}\label{r3.12}
x\Sigma (x,Q^2)\sim\rho^{-1}\sigma^{-1/2}e^{2\gamma\sigma -\delta
\zeta}(1+\sum\limits_{i=1}^n\varepsilon_f^i\rho^{i+1}\alpha_s^i)
\end{equation}
where $\xi$, $\zeta$, $\sigma$, $\rho$ are given as:
\begin{equation}\label{r3.13}
\xi =\ln (\frac{x_0}{x});~~\zeta =\ln\left(\frac{\alpha_s(Q_0^2)}
{\alpha_s(Q^2)}\right);~~\sigma =\sqrt{\xi\zeta};
~~\rho =\sqrt{\xi /\zeta}
\end{equation}
$n=0$  in the summation (\ref{r3.11}) and (\ref{r3.12}) corresponds to
LO while $n=1$ to NLO approximations. Remaining parton distributions
{\it i.e.} $p=q_{NS}$, $\Delta q_{NS}$, $\Delta\Sigma$, $\Delta g$
behaves in the small $x$ region like:
\begin{equation}\label{r3.14}
p(x,Q^2)\sim \sigma^{-1/2}e^{2\gamma_f\sigma -\delta_f\zeta}
(1+\sum\limits_{i=1}^n\varepsilon_f^i\rho^{2i+1}\alpha_s^i)
\end{equation}
One can read from (\ref{r3.11}), (\ref{r3.12}) and (\ref{r3.14}), that
spin dependent parton distribution functions $x\Delta q_{NS}$,
$x\Delta\Sigma$, $x\Delta g$ and the unpolarised nonsinglet quark
distribution $xq_{NS}$ are by a factor of $x$ less singular than the
unpolarised distribution $xg$ and $x\Sigma$. Moreover, in the case of
$p$ function ($p=q_{NS}$, $\Delta q_{NS}$, $\Delta\Sigma$, $\Delta g$)
(\ref{r3.14}) higher order corrections are more important than in the
$xg$ (\ref{r3.11}) and $x\Sigma$ (\ref{r3.12}) case because of
appearance $\rho^{2i+1}$ terms in (\ref{r3.14}) instead of
$\rho^{i+1}$ in (\ref{r3.11}) and (\ref{r3.12}). It must be also
emphasised, that the small $x$ behaviour of parton distributions
strongly depends on the input parametrisation $p(x,Q_0^2)$
(\ref{r2.32}). When it is nonsingular, then the singular small $x$
behaviour of parton distributions is fully generated by $Q^2$
evolution and has a form (\ref{r3.11})-(\ref{r3.14}). Whereas in
a case of singular input parametrisation {\it e.g.}\cite{b7}
\begin{equation}\label{r3.15}
\Delta g(x,Q_0^2)\sim x^{\alpha}(1-x)^{\beta}(1+\gamma x^{\delta})
\end{equation}
where $\alpha =-0.5$, $\beta =4$, $\gamma =3$, $\delta =1$, singular
$x^{\alpha}$ small $x$ behaviour of $\Delta g$ distribution will
survive $Q^2$ QCD evolution and will be leading towards the singular
terms, generated perturbatively. Choosing the input parametrisation
(\ref{r2.32}), experimental data must be taken into account. As a
start scale for the GLAP evolution equations $Q_0^2=1~{\rm GeV}^2$ is
assumed at present. This value is a limit where from the one side
($Q^2<Q_0^2$) experimental measurements confirm the Regge theory of
DIS, and from the other side $Q^2>Q_0^2$ it is a starting point for
methods of perturbative QCD ($Q^2\ge 1~{\rm GeV}^2$). There are two
approaches dependent on the choice of the input parametrisation.
Either we assume Regge behaviour (\ref{r3.2}) of parton distributions
at small $x$ and then their shape  at small $x$ and larger $Q^2$
$Q^2>Q_0^2$ is fully implied by NLO QCD evolution
(\ref{r3.11})-(\ref{r3.14}) or we take singular input parametrisations
(but no more singular than it results from experimental data
{\it i.e.} $\Delta q_{NS}(x,Q^2)\le x^{-0.5}$), which will survive LO
and NLO GLAP evolution. The Regge theory predicts the following
behaviour of parton distributions at small $x$ and
$Q^2\le 1~{\rm GeV}^2$:
\begin{eqnarray}\label{r3.16}
x\Sigma&\sim& const~(Pomeron)\nonumber\\
q_{NS}&\sim& x^{-0.5}~(Reggeon~A_2:\rho -\omega);\nonumber\\
\Delta\Sigma,\Delta q_{NS}&\sim& x^0\div x^{0.5}~(Reggeon~A_1)
\end{eqnarray}
Experimental analyses \cite{b5},\cite{b19} are in a good agreement
with (\ref{r3.16}) for $Q^2\le 1~{\rm GeV}^2$ and also for larger
$Q^2$, after taking into account the perturbative effects.
Investigating small $x$ region within perturbative methods, one should
include all of those terms in $C(x,Q^2)$ and $P_{ij}(x,\alpha_s)$,
which remarkably influence the shape of the nucleon structure
functions and what will be soon verified experimentally. It has been
lately noticed \cite{b20},\cite{b24},\cite{b25} that the spin
dependent structure function $g_1$ in the small $x$ region is
dominated by $\ln^2(1/x)$ terms. These contributions correspond to the
ladder diagrams with quark  and gluon exchanges along the ladder -
{\it cf} Fig.2. The contribution of non-ladder diagrams to the
nonsinglet spin dependent structure function is negligible.
\begin{figure}[ht]
\begin{center}
\includegraphics[width=80mm]{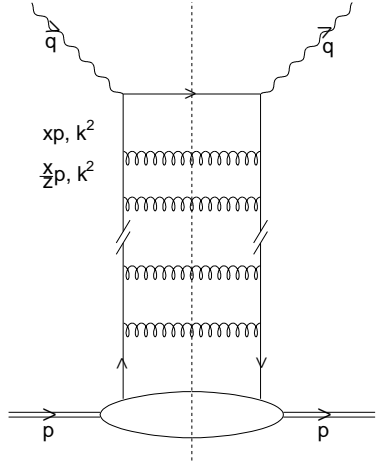}
\caption{A ladder diagram generating double logarithmic $\ln^2(1/x)$
terms in the nonsingled spin structure function $g_1$.}
\end{center}
\end{figure}
Thus the behaviour of the spin dependent nucleon structure functions
at small $x$ is expected to be governed by leading double logarithmic
terms of type $\alpha_s^n\ln^{2n}(x)$. These terms must be resummed
in the coefficients  and splitting functions $P_{ij}(x,\alpha_s^2)$.
The resummation of the double logarithmic terms $\ln^2x$ in the limit
of a very small $x$ ($x\rightarrow 0$) is given by the following
equations \cite{b20}:
\begin{equation}\label{r3.17}
f_{NS}(x,k^2)=f_{NS}^{(0)}(x,k^2)+\frac{\alpha_s}{2\pi}
\int\limits_x^1\frac{dz}{z}\int\limits_{k_0^2}^{k^2/z}
\frac{dk'^2}{k'^2}\Delta P_{qq}^{(0)}(z)f_{NS}(\frac{x}{z},k'^2)
\end{equation}
\begin{eqnarray}\label{r3.18}
f_S(x,k^2)=f_S^{(0)}(x,k^2)+\frac{\alpha_s}{2\pi}
\int\limits_x^1\frac{dz}{z}\int\limits_{k_0^2}^{k^2/z}
\frac{dk'^2}{k'^2}\nonumber\\
\times\left[\Delta P_{qq}^{(0)}(z)f_S(\frac{x}{z},k'^2)
+\Delta P_{qG}^{(0)}(z)f_g(\frac{x}{z},k'^2)\right]
\end{eqnarray}
\begin{eqnarray}\label{r3.19}
f_g(x,k^2)=f_g^{(0)}(x,k^2)+\frac{\alpha_s}{2\pi}
\int\limits_x^1\frac{dz}{z}\int\limits_{k_0^2}^{k^2/z}
\frac{dk'^2}{k'^2}\nonumber\\
\times\left[\Delta P_{Gq}^{(0)}(z)f_S(\frac{x}{z},k'^2)
+\Delta P_{GG}^{(0)}(z)f_g(\frac{x}{z},k'^2)\right]
\end{eqnarray}
where $\Delta P_{ij}^{(0)}(z)$ are LO approximation at
$z\rightarrow 0$ and have a form (\ref{r3.7}). Unintegrated
distributions $f$ in equations (\ref{r3.17})-(\ref{r3.19}) are related
to the corresponding polarised distributions $\Delta p(x,Q^2)$ via
\begin{equation}\label{r3.20}
\Delta p(x,Q^2)=\Delta p^{(0)}(x)+\int\limits_{k_0^2}^{W^2}
\frac{dk^2}{k^2}f(x'=x(1+k^2/Q^2),k^2)
\end{equation}
\begin{equation}\label{r3.21}
\Delta p^{(0)}(x)=\int\limits_0^{k_0^2}\frac{dk^2}{k^2}f(x,k^2)
\end{equation}
$k^2$ is the transverse momentum squared of the parton. The
inhomogeneous term $f^{(0)}(x,k^2)$ in the nonsinglet case
(\ref{r3.17}) has a form
\begin{eqnarray}\label{r3.22}
f_{NS}^{(0)}(x,k^2)=\frac{\alpha_s(k^2)}{2\pi}\frac43
\int\limits_x^1\frac{dz}{z}\frac{(1+z^2)\Delta p^{(0)}(x/z)
-2z\Delta p^{(0)}(x)}{1-z}\nonumber\\
+\frac{\alpha_s(k^2)}{2\pi}[2+
\frac83\ln (1-x)]\Delta p^{(0)}(x)
\end{eqnarray}
where the nonperturbative part of the distribution $\Delta p^{(0)}(x)$
is parametrised on the basis of the experimental data at the small
values of $Q^2$ ($Q^2<1~{\rm GeV}^2$). This nonperturbative
parametrisation is given by
\begin{equation}\label{r3.23}
\Delta p_{NS}^{(0)}(x)=N(1-x)^\eta
\end{equation}
Parameter $\eta$ in the valence quark case ($\Delta q_{NS}$) is equal
to 3, $g_A=1.257$ is the axial vector coupling and $N$ is the
normalisation constant, in the nonsinglet case determined from the
Bjorken sum rule (\ref{r2.11}), (\ref{r2.12}), which can be written
as:
\begin{equation}\label{r3.24}
\int\limits_0^1 g_1^{NS}(x,Q_0^2)dx=
\int\limits_0^1 (g_1^p-g_1^n)(x,Q_0^2)dx=\frac16g_A
\end{equation}
The source of the double logarithmic terms $\ln^2x$ in $g_1(x,Q^2)$ is
the double integration in the formula for function $f(x,k^2)$:
\begin{equation}\label{r3.25}
f(x,k^2)\sim\frac{\alpha_s}{2\pi}\int\limits_x^1\frac{dz}{z}
\int\limits_{k_0^2}^{k^2/z}\frac{dk'^2}{k'^2}
\end{equation}
where the upper limit in the integral over the transverse momentum
$k'^2$ is $z$-dependent  ($=k^2/z$). Thus, double logarithmic terms
come from the integration over the longitudinal momentum fraction $z$
together with the integration over $k'^2$ with $z$-dependent upper
limit:
\begin{equation}\label{r3.26}
f(x,k^2)\sim\ln^2(1/x)=\ln^2x
\end{equation}
Equations (\ref{r3.17})-(\ref{r3.19}) in the case of the fixed
coupling constant $\alpha_s$ can be solved analytically \cite{b20}. These
equations generate singular small $x$ behaviour of the polarised
parton distributions and hence of the spin dependent structure
function $g_1$ {\it i.e.}:
\begin{equation}\label{r3.27}
g_1^{NS}(x,Q^2)\sim x^{-\lambda_{NS}};~~g_1^S(x,Q^2)
\sim x^{-\lambda_S};~~\Delta g(x,Q^2)\sim x^{-\lambda_S}
\end{equation}
where $g_1^{NS}=g_1^p-g_1^n$, $g_1^S=g_1^p+g_1^n$, $\Delta g$ is the
polarised gluon distribution and exponents $\lambda_i$ have forms:
\begin{equation}\label{r3.28}
\lambda_{NS}=2\sqrt{\frac{\alpha_s}{2\pi}\Delta P_{qq}^{(0)}(x)};~~~
\lambda_S=2\sqrt{\frac{\alpha_s}{2\pi}\gamma^+}
\end{equation}
\begin{eqnarray}\label{r3.29}
\gamma^+&=&\frac12[\Delta P_{qq}^{(0)}(x)+\Delta P_{GG}^{(0)}(x)\nonumber\\
&+&\sqrt{[\Delta P_{qq}^{(0)}(x)-\Delta P_{GG}^{(0)}(x)]^2
+4\Delta P_{qG}^{(0)}(x)\Delta P_{Gq}^{(0)}(x)}]
\end{eqnarray}
As it has already been mentioned above, the singular small $x$
behaviour of the polarised structure function (\ref{r3.27}) become
leading only in the case of the nonsingular input parametrisation
(\ref{r2.32}) {\it e.g.} for the simple parametrisation (\ref{r3.23}).
Because only unpolarised, nonsinglet parton distributions $q_{NS}$
have Regge small $x$ behaviour $x^{-0.5}$  more singular than that,
implied by QCD evolution, the shape of all spin dependent
distributions is mostly governed by QCD evolution. Thus the leading
small $x$ behaviour of polarised nucleon structure functions is
(\ref{r3.27}) and moreover singlet structure functions dominate over
the nonsinglet ones ($\lambda_S>\lambda_{NS}$). For example, at
$Q^2=10~{\rm GeV}^2$, $N_f=3$ and scale parameter $\Lambda=232 MeV$, the
corresponding values of $\lambda_i$ are: $\lambda_S=1.4$,
$\lambda_{NS}=0.48$ ($\alpha_s=0.27$, $\gamma^+=11.7$). It must be
pointed out, that the equations (\ref{r3.17})-(\ref{r3.19}) can be
applied only for very small $x$. They are based on the approximation,
that for very small $x$ splitting functions $\Delta P_{ij}^{(0)}(x
\rightarrow 0)$ are given by (\ref{r3.7}). At large and moderately
small values of $x$ this approach is no more adequate. In the region
of "not too small $x$", which can be explored experimentally, in
theoretical QCD analyses one should use the GLAP (LO or NLO) equations
with complete $\Delta P_{ij}(z)$ functions. Combining the standard LO
GLAP approach with the double $\ln^2x$ resummation, it is possible on
the one hand to guarantee an agreement of QCD predictions with
experimental data in the large and moderately small $x$ and on the
other hand to generate the singular small $x$ shape of polarised
structure functions, governed by $\ln^2x$ terms. With this aim the
equations (\ref{r3.17})-(\ref{r3.19}) should be extended to include
complete splitting functions $\Delta P_{ij}(z)$ and not only their
approximations (\ref{r3.7}) for $z\rightarrow 0$. In this way one can
obtain system of equations, containing both LO GLAP evolution and the
double logarithmic $\ln^2x$ effects at small $x$. Analyses of such
unified GLAP LO + $\ln^2x$ approach are presented in \cite{b20}. On
the basis of this interesting method we give in the next chapter the
predictions for the $g_1^{NS}$ function in the case of nonsingular as
well as singular input parametrisation $\Delta p_{NS}(x,Q_0^2)$.

\section{Predictions for the nonsinglet spin structure function $g_1$}

The small $x$ behaviour of both nonsinglet and singlet spin dependent
structure functions $g_1^{NS}(x,Q^2)$ and $g_1^S(x,Q^2)$ is governed
by the double logarithmic terms $\alpha_s^n\ln^{2n}(x)$ \cite{b20},
\cite{b24},\cite{b25}. But in contrast to the singlet polarised
function, for the nonsinglet one the contribution of nonladder
diagrams is negligible. Thus we should consider only ladder diagrams
with quark (antiquark) exchange, Fig.2. Hence the nonsinglet part of
the polarised structure function $g_1$ has a form:
\begin{equation}\label{r4.1}
g_1^{NS}(x,Q^2)=g_1^p(x,Q^2)-g_1^n(x,Q^2)
\end{equation}
where $g_1^p$ and $g_1^n$ are spin dependent structure functions of
proton and neutron respectively. According to (\ref{r2.3}),
(\ref{r2.4}), (\ref{r2.29})-(\ref{r2.31}) one can obtain for the
colour number $N_c=3$:
\begin{equation}\label{r4.2}
g_1^p=\frac29\Delta u+\frac{1}{18}\Delta d+\frac{5}{18}\Delta\bar{u}+
\frac19\Delta\bar{s}
\end{equation}
\begin{equation}\label{r4.3}
g_1^n=\frac{1}{18}\Delta u+\frac29\Delta d+\frac{5}{18}\Delta\bar{u}+
\frac19\Delta\bar{s}
\end{equation}
and hence
\begin{equation}\label{r4.4}
g_1^{NS}=\frac16(\Delta u_{val}-\Delta d_{val})=\frac16(\Delta u-\Delta d)
\end{equation}
The simple form of $g_1^{NS}$ (\ref{r4.4}) results from the assumption
of SU(3) flavour symmetry:
\begin{equation}\label{r4.5}
\Delta\bar{u}=\Delta\bar{d}
\end{equation}
and hence all of gluon and sea quark contributions from the proton and
the neutron structure function cancel mutually. This feature that the
small $x$ behaviour of the spin dependent nonsinglet structure
function is governed by the double logarithmic terms
$\alpha_s^n\ln^{2n}(x)$ is very important from the point of view of
small $x$ QCD analysis. This is different from the case of unpolarised
nonsinglet structure functions $F_2^{NS}$, where the small $x$
behaviour of $F_2$, generated by the $\alpha_s^n\ln^{2n}(x)$ terms, is
dominated by the nonperturbative contribution of $A_2$ Regge pole. For
$g_1^{NS}$ the relevant $A_1$ Regge pole has low intercept
$\alpha_{NS}(0)\le 0$ and for small $x$ in the Regge limit one has:
\begin{equation}\label{r4.6}
g_1^{NS}(x,Q^2)\sim x^{-\alpha_{NS}(0)}
\end{equation}
Thus the Regge behaviour of the spin dependent structure functions is
unstable against the resummation of the $\ln^2x$ terms, which generate
more singular $x$ shape than that (\ref{r4.6}) with
$\alpha_{NS}(0)\le 0$. Therefore the measurement of the nonsinglet
spin dependent structure function can be a very important test of the
QCD perturbative analyses in the small $x$ region. In our numerical
analysis we follow \cite{b20} and \cite{b25}. Solving the unified
equation incorporating GLAP $Q^2$ evolution and the $\ln^2x$
resummation we get the results for the nonsinglet polarised structure
function $g_1^{NS}(x,Q^2)$ in the perturbative region $Q^2>Q_0^2$ for
different values of $x\in(0;1)$. This equation taking into account
both GLAP evolution and $\ln^2x$ effects for $g_1^{NS}$ function has
a form \cite{b20},\cite{b25}:
\begin{eqnarray}\label{r4.7}
f(x,k^2)&=&f^{(0)}(x,k^2)+\frac{2\alpha_s(k^2)}{3\pi}
\int\limits_x^1\frac{dz}{z}\int\limits_{k_0^2}^{k^2/z}
\frac{dk'^2}{k'^2}f(\frac{x}{z},k'^2)\nonumber\\
&+&\frac{\alpha_s(k^2)}{2\pi}
\int\limits_{k_0^2}^{k^2}\frac{dk'^2}{k'^2}[\frac43
\int\limits_x^1\frac{dz}{z}\frac{(z+z^2)f(x/z,k'^2)-2zf(x,k'^2)}{1-z}
\nonumber\\
&+&(\frac12+\frac83\ln (1-x))f(x,k'^2)]
\end{eqnarray}
where
\begin{eqnarray}\label{r4.8}
f^{(0)}(x,k^2)&=&\frac{\alpha_s(k^2)}{2\pi}
[\frac43\int\limits_x^1\frac{dz}{z}\frac{(1+z^2)g_1^{(0)}(x/z)
-2zg_1^{(0)}(x)}{1-z}\nonumber\\
&+&(\frac12+\frac83\ln (1-x))g_1^{(0)}(x)]
\end{eqnarray}
\begin{equation}\label{r4.9}
g_1(x,Q^2)=g_1^{(0)}(x)+\int\limits_{k_0^2}^{Q^2(1/x-1)}
\frac{dk^2}{k^2}f\left(x(1+\frac{k^2}{Q^2}),k^2\right)
\end{equation}
and
\begin{equation}\label{r4.10}
g_1^{(0)}(x)=\int\limits_0^{k_0^2}\frac{dk^2}{k^2}f(x,k^2)
\end{equation}
Comparing (\ref{r4.7})-(\ref{r4.10}) with (\ref{r3.17}),
(\ref{r3.20})-(\ref{r3.22}) it is clear that $\Delta P_{qq}(x)$ in
(\ref{r4.7}) has a full GLAP form instead of its approximation for
$x\rightarrow 0$ in (\ref{r3.17}) and $g_1^{NS}$ plays simply  the
role of $\Delta p$ from (\ref{r3.20}). We solve eq.(\ref{r4.7}) using
different parametrisations of $g_1^{NS(0)}(x)$: the simple one,
implied by Regge behaviour of $g_1^{NS}$ in nonperturbative region
\begin{equation}\label{r4.11}
g_1^{NS(0)}(x)\equiv g_1^{NS}(x,Q_0^2)=N(1-x)^3
\end{equation}
and two dynamical inputs: GRSV (Gl\"uck, Reya, Stratmann, Vogelsang)
\cite{b22} and GS (Gehrmann, Stirling) \cite{b23}. The nonsinglet spin
dependent structure function must satisfy the Bjorken sum rule
(\ref{r2.11}), (\ref{r2.12}) independently of the value of $Q^2$. This
means that for any $Q^2$, the first moment of $g_1^{NS}$ must be equal
to $1/6 g_A$ similarly to the case of the low scale $Q_0^2$
(\ref{r3.24}):
\begin{equation}\label{r4.12}
\langle g_1^{NS}(x,Q^2)\rangle\equiv\int\limits_0^1g_1^{NS}(x,Q^2)dx=
\int\limits_0^1(g_1^p-g_1^n)(x,Q^2)dx=\frac16g_A=0.2095
\end{equation}
This condition implies the proper normalisation constants $N$ in all
of input parametrisations. Thus the constant $N$ in (\ref{r4.11}),
found from the Bjorken sum rule is equal to $2/3 g_A=0.838$ (we set
$g_A=1.257$) and the Regge nonsingular input (\ref{r4.11}) takes a
form:
\begin{equation}\label{r4.13}
REGGE:~~~g_1^{NS}(x,Q_0^2)=\frac23g_A(1-x)^3=0.838(1-x)^3
\end{equation}
The Regge behaviour of structure functions at small $x$, as it was
mentioned above, has been confirmed by HERA experiments in the low
$Q^2$ region ($Q^2<1~{\rm GeV}^2$). Therefore the choice of the Regge
input allows to unite the nonperturbative origin with QCD
perturbative analysis starting at $Q_0^2\sim 1~{\rm GeV}^2$. In this
way, assuming the Regge (flat, nonsingular) behaviour of structure
functions at low $Q^2$ scale {\it i.e.} $Q_0^2=1~{\rm GeV}^2$, we
expect that the singular small $x$ behaviour of polarised structure
functions is completely generated by QCD evolution, involving NLO or
even (as in our case) GLAP+$\ln^2x$ approach. This analysis, based on
the Regge input (\ref{r4.13}), is however one of two main possible
scenarios, describing the small $x$ behaviour of spin structure
functions. The second is to allow steeper (more singular) inputs of
structure functions at $Q_0^2$, what intensifies more the growth of
structure functions as $x\rightarrow 0$ implied by QCD. The only
constraint on these two scenarios is consistency of their predictions
with experimental data. In our analysis of the $g_1^{NS}$ structure
function we consider dynamical inputs proposed by GRSV \cite{b22} and
GS \cite{b23}. These inputs result from a global analysis of all
available recently deep inelastic polarised structure function data
\cite{b14}. Our calculations incorporating both GLAP evolution and
resummation of the $\ln^2x$ terms are based on the LO fitted inputs.
In such a way the spin dependent nonsinglet structure function
$g_1^{NS}$ (\ref{r4.4}) has an input form:
\begin{eqnarray}\label{r4.14}
GRSV&:&g_1^{NS}(x,Q_0^2=1~{\rm GeV}^2)=0.327x^{-0.267}
(1-0.583x^{0.175}+1.723x\nonumber\\
&+&3.436x^{3/2})
(1-x)^{3.486}+0.027x^{-0.624}(1+1.195x^{0.529}\nonumber\\
&+&6.164x+2.726x^{3/2})(1-x)^{4.215}
\end{eqnarray}
\begin{eqnarray}\label{r4.15}
GS&:&g_1^{NS}(x,Q_0^2=4~{\rm GeV}^2)=0.29x^{-0.422}
(1+9.38x-4.26\sqrt{x})~~~~~~~~\nonumber\\
&\times&(1-x)^{3.73}+
0.196x^{-0.334}(1+10.46x-5.10\sqrt{x})(1-x)^{4.73}
\end{eqnarray}
for details see Appendix B. All numerical calculations have been
performed in C code on PC computer under LINUX system. Our numerical
results for $g_1^{NS}$ based on Regge (\ref{r4.13}), GRSV
(\ref{r4.14}) and GS (\ref{r4.15}) input parametrisations are
presented in Figs.3-7. In Fig.3 we plot different input
parametrisations $g_1^{NS}(x,Q_0^2)$. Figs.4,5 show the nonsinglet
function $g_1^{NS}$ after evolution to $Q^2=10~{\rm GeV}^2$ for these
different parametrisations (Regge, GRSV, GS) and Figs.6,7 present the
function $6xg_1^{NS}=x(\Delta u_{val}-\Delta d_{val})$ at
$Q^2=10~{\rm GeV}^2$ also for different inputs $g_1^{NS}(x,Q_0^2)$.
In all of Figs.4-7 pure GLAP evolution is compared with double
logarithmic $\ln^2x$ effects at small $x$. Additionally, in Figs.5-7
we compare our numerical results with recent SMC (1997) data
\cite{b14}. Contributions $6\langle g_1^{NS}\rangle$ (\ref{r4.12})
and $6\Delta I(x_a,x_b,Q^2)$
\begin{equation}\label{r4.16}
\Delta I(x_a,x_b,Q^2)\equiv\int\limits_{x_a}^{x_b}g_1^{NS}(x,Q^2)dx
\end{equation}
to the Bjorken sum rule at $Q^2=10~{\rm GeV}^2$ together with
experimental SMC values are presented in Tab.2: *) means the
extrapolation of experimental data to low $x$ and **) is the integral
over the measured range of $x$.
\begin{table}[ht]
\title{TABLE II}
\begin{center}
\begin{tabular}{|c|c|c|c|}
\hline
PARAMETRISATION & $6\Delta I$ & $6\Delta I$ &
$6\Delta I$\\
 & $(0,1,Q^2)$ & $(0,0.003,Q^2)$ & $(0.003,0.7,Q^2)$\\ \hline
INPUT & 1.257 & 0.0150 & 1.232\\
{\bf REGGE} LO GLAP & 1.255 & 0.0342 & 1.219\\
LO GLAP+$\ln^2x$ & 1.249 & 0.0493 & 1.198\\ \hline
INPUT & 1.257 & 0.0786 & 1.194\\
{\bf GRSV} LO GLAP & 1.249 & 0.107 & 1.171\\
LO GLAP+$\ln^2x$ & 1.242 & 0.119 & 1.153\\ \hline
INPUT & 1.257 & 0.123 & 1.160\\
{\bf GS} LO GLAP &  1.253 & 0.134 & 1.151\\
LO GLAP+$\ln^2x$ & 1.247 & 0.142 & 1.139\\ \hline
EXPERIMENT & 1.29$\pm$0.24 & *) 0.09$\pm$0.09 & **) 1.20$\pm$0.24\\ \hline
\end{tabular}
\caption{Theoretical contributions $6\Delta I(x_a,x_b,Q^2)$
and their experimental SMC values}
\end{center}
\end{table}
From Figs.4-7 one can read that the double logarithmic $\ln^2x$
effects are very significant for $x\le 10^{-2}$. Besides, as it has
been expected, the growth of the nonsinglet proton spin structure
function $g_1^{NS}$ in the very small $x$ region is much steeper for
dynamical parametrisations (GRSV or GS) than for the Regge one. The
comparison of our theoretical model with experimental data in Tab.2
and Figs.5-7 yields the conclusion that all of the theoretical
predictions for different parametrisations (Regge, GRSV, GS) and
incorporating pure LO GLAP QCD evolution as well as LO GLAP evolution
with $\ln^2x$ effects are in a good agreement with experimental data
within statistical errors. Unfortunately, the most interesting $x$
region is still nonavailable for experiment. So the problem, which QCD
approach is the most adequate for the description of small $x$ physics
in the polarised deep-inelastic scattering of particles remains
unsolved.
\begin{figure}[ht]
\begin{center}
\includegraphics[width=100mm]{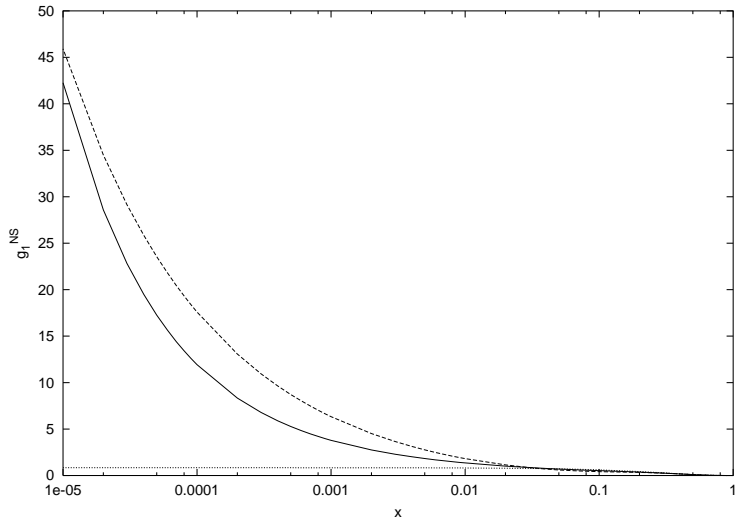}
\caption{Input parametrisations of the nonsingled spin structure
function of the proton $g_1^{NS}(x,Q_0^2)$: REGEE (\ref{r4.13})
- dotted line; GRSV (\ref{r4.14}) - solid line; GS (\ref{r4.15})
- dashed line.}
\end{center}
\end{figure}
\clearpage

\begin{figure}[ht]
\begin{center}
\includegraphics[width=100mm]{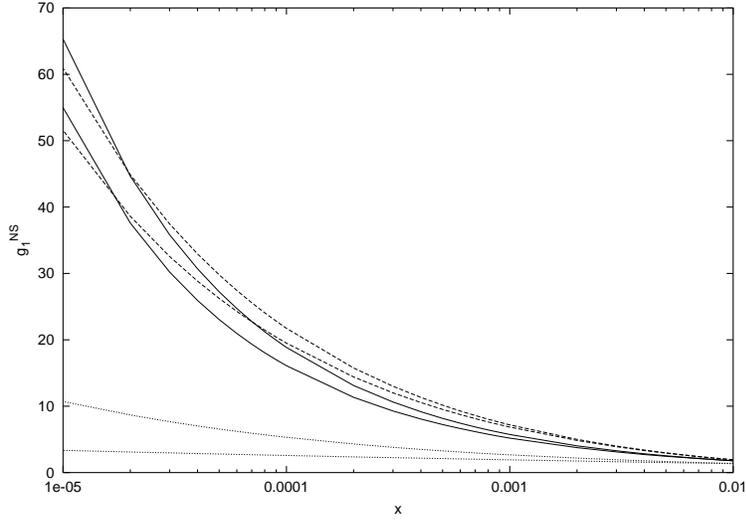}
\caption{$g_1^{NS}$ at $Q^2=10~{\rm GeV}^2$ based on inputs: REGGE - dotted,
GRSV - solid, GS - dashed. For each pair of lines the pure LO GLAP
prediction lies below the LO GLAP+$\ln^2x$ one.}
\end{center}
\end{figure}

\begin{figure}[h]
\begin{center}
\includegraphics[width=100mm]{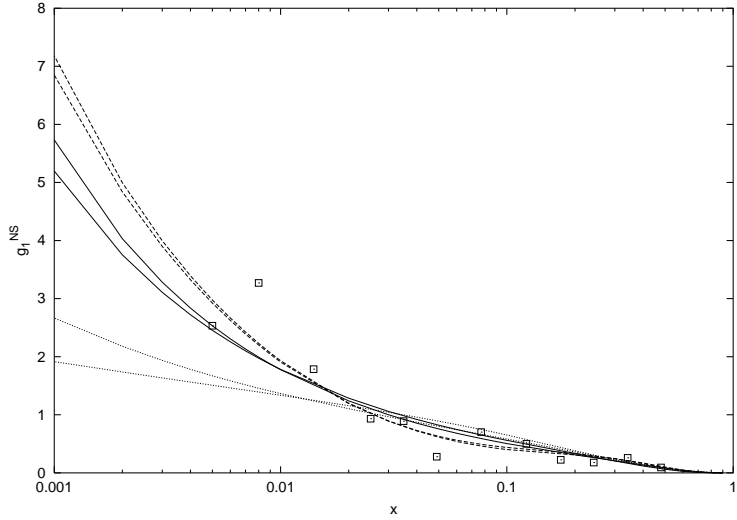}
\caption{$g_1^{NS}$ at $Q^2=10~{\rm GeV}^2$; similarly as in Fig.4 but for
the measurable experimentally region of $x$. Squares show the recent SMC
data 1997 \cite{b14}.}
\end{center}
\end{figure}
\clearpage

\begin{figure}[ht]
\begin{center}
\includegraphics[width=98mm]{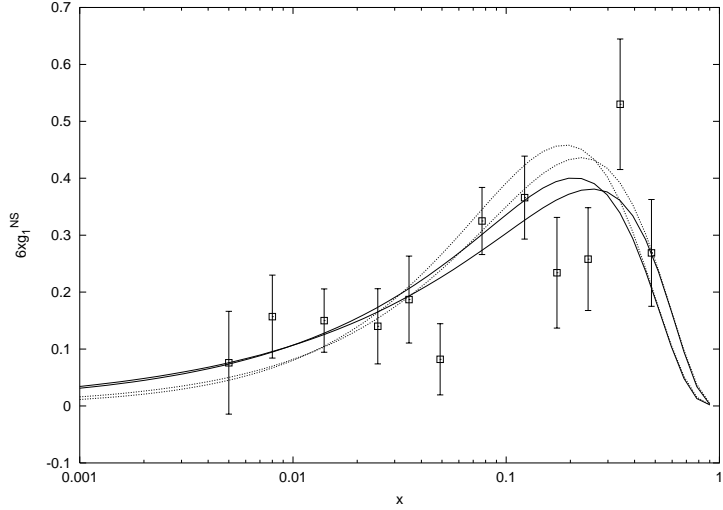}
\caption{Function $6xg_1^{NS}$ at $Q^2=10~{\rm GeV}^2$. Predictions
based on the REGGE input - dotted and GRSV - solid. LO GLAP above LO
GLAP+$\ln^2x$ at $x=0.2$. SMC 1997 data with statistical errors are shown.}
\end{center}
\end{figure}

\begin{figure}[h]
\begin{center}
\includegraphics[width=98mm]{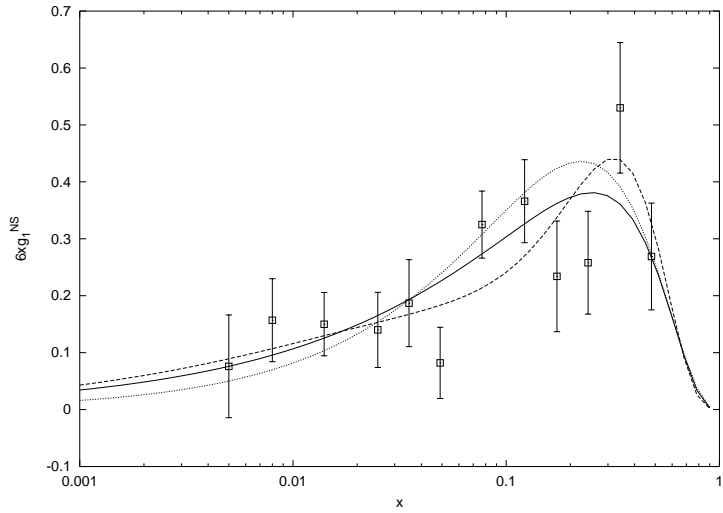}
\caption{LO GLAP+$\ln^2x$ predictions for function $6xg_1^{NS}$ at
$Q^2=10~{\rm GeV}^2$, based on input parametrisations: REGGE - dotted,
GRSV - solid, GS - dashed. Plots are compared with SMC 1997 data.}
\end{center}
\end{figure}
\pagebreak

\section{Summary and conclusions}

In this paper the main theoretical and experimental problems in
nucleon spin structure physics have been briefly reviewed. The results
of current experiments are deviation of the Ellis-Jaffe sum rule and
validity of the Bjorken sum rule. This causes that the question "how
is the spin of the nucleon made out of partons?" is still open. The
great puzzle are experimental results which violating the Ellis-Jaffe
sum rule imply that only a very small part of the spin of the proton
is carried by quarks. So where is the nucleon spin? Maybe gluons take
a large fraction of the nucleon spin? Or maybe the spin of the proton
is "hidden" in orbital angular momentum of quarks and gluons? Maybe at
last the solution of the spin problem lies in the small $x$ physics
and the lacking spin of the nucleon is hidden in the unmeasured very
small $x$ region. The answer the above questions will be possible
thanks to the progress in theoretical and experimental research in the
small $x$ physics. Perturbative QCD analysis, based on GLAP evolution
equations is in a good agreement with experimental data. This
agreement concern unpolarised and polarised structure functions of the
nucleon $F_1$, $F_2$, $g_1$ within NLO approximation in the large and
the moderately small Bjorken $x$ region. Unfortunately, practically
lack of the experimental measurements in the very small $x$ region
($x\le 10^{-3}$) makes the satisfactory verification of the
theoretical QCD predictions in this region impossible. Knowledge of
the behaviour of the nucleon spin structure functions when
$x\rightarrow 0$ is crucial in the determination of Bjorken and
Ellis-Jaffe sum rules {\it i.e.} in overcoming the "spin crisis".
Understanding of the small $x$ physics in the polarised DIS processes
requires to take into account all of these perturbative QCD effects
which become significant in the small $x$ region and which could be
verified by future experiments. Present QCD analyses, based on the
GLAP LO or NLO $Q^2$ evolution seem to be incomplete when
$x\rightarrow 0$. The growth of the unpolarised as well polarised
structure functions of the nucleon in the small $x$ region is governed
by leading double logarithmic terms of the form
$\alpha_s^n\ln^{2n}(x)$, generated by ladder diagrams with quark and
gluon exchange. This singular behaviour of the structure functions at
low $x$, implied by $\ln^2x$ terms, is however better visible in the
polarised case. For unpolarised, nonsinglet structure functions of the
nucleon the QCD evolution behaviour at small $x$  is screened by the
leading Regge contribution. Therefore the spin dependent structure
functions of the nucleon are a sensitive test of the perturbative QCD
analyses in the low $x$ region. Our numerical analyses incorporating
the LO GLAP evolution and the $\ln^2x$ effects at small $x$ show, that
the growth of the nonsinglet polarised structure function of the
nucleon $g_1^{NS}$, implied by $\ln^2x$ terms, is significant for
$x\le 10^{-2}$. Our predictions for $g_1^{NS}$ are in a good agreement
with the recent SMC data for small $x$ region ($x\sim 10^{-3}$). The
contribution from the low $x$ region ($x\le 0.003$) to the Bjorken sum
rule is found to be around 4\% (for Regge input $g_1^{NS}(x,Q_0^2)$)
and 10\% (for dynamical inputs) of the value of the sum. Theoretical
predictions for $g_1^{p,n,d}$, taking into account the $\ln^2x$
resummation effects will be in the future verified  experimentally.
There are a few hopeful experimental projects of the investigation of
the nucleon's spin structure. One of these is the HERMES experiment
(start in 1995) located in HERA at DESY with a fixed polarised H,D or
$^3{\rm He}$ target and a longitudinally polarised positron beam of
27.5 GeV \cite{b26}. The accessible kinematic range is $0.004<x<1$ and
$0.2<Q^2<20~{\rm GeV}^2$. The HERMES experiment allows a direct
measurement of the polarised quark distributions for individual
flavours also $g_1^{p,n,d}(x,Q^2)$ and even $g_2(x,Q^2)$. The question
of the gluon polarisation is also addressed experimentally. The
polarised gluon distribution $\Delta g(x,Q^2)$ may play a crucial role
in understanding of the nucleon spin structure. The measurement of
$\Delta g(x,Q^2)$ in the charm production via photon-gluon fusion
process $\gamma^*g\rightarrow c\bar{c}$ will be possible at COMPASS
experiment at CERN \cite{b15}. In this project the polarised muons
will be scattered on polarised proton and deuteron targets. The energy
of the muon beam will be of 100~GeV and 200~GeV and the Bjorken $x$
region $x>0.02$. The COMPASS measurements are expected to start in
2000. A very important program which will test many elements of QCD
in the perturbative as well as in the nonperturbative region is RHIC
spin project at Brookhaven \cite{b21}. This program with polarised
proton-proton collider will start in 2000 and will allow for a
measurement of the polarised gluon density via heavy quark production
($gg\rightarrow Q\bar{Q}$) or via direct photon production
($gq\rightarrow\gamma q$). Finally, a very promising experimental
project in high energy spin physics is HERA \cite{b8}. The
polarisation of the proton and electron beams at
$\sqrt{s}=300~{\rm GeV}$ will enable to measure the structure function
$g_1(x,Q^2)$ and spin dependent quark distributions
$\Delta q_f(x,Q^2)$ at very low $x$ ($x\sim 10^{-5}$). From polarised
di-jet production it will be possible to determinate the polarised
gluon distribution $\Delta g(x,Q^2)$ for the region $0.002<x<0.2$.
Additionally in HERA, a program of polarised proton-proton collisions
is proposed. This high energy proton-proton scattering will allow via
$J/\psi$ production for the direct determination of the gluon function
$\Delta g(x,Q^2)$. The new HERA with polarised experiments and the
largely extended kinematical region of $x$ and $Q^2$ will contribute
a lot to our understanding of high energy spin physics. The problem of
the spin structure of the nucleon is nowadays one of the most
important challenges for theory and experiment as well.

\section*{Acknowledgements}

I would like to thank Jan Kwieci\'nski for a great help and critical
remarks during preparing this work. I am also grateful to Andrzej
Kotlorz for useful discussions about numerical problems.

\appendix
\section{GLAP evolution of polarised quark and gluon distribution
functions in the nucleon}

In perturbative QCD distribution functions of partons $q(x,Q^2)$,
$g(x,Q^2)$, $\Delta q(x,Q^2)$, $\Delta g(x,Q^2)$ evolve with $Q^2$.
This evolution is described by GLAP \cite{b3},\cite{b9} equations and
it is assumed that for polarised functions $\Delta p(x,Q^2)$ the
evolution equations have the same form like for unpolarised functions
$p(x,Q^2)$:
\begin{equation}\label{rA.1}
\frac{d}{dt}[\Delta q_{NS}(x,t)]=\frac{\alpha_s(t)}{2\pi}
\Delta P_{qq}\otimes\Delta q_{NS}(x,t)
\end{equation}
\begin{equation}\label{rA.2}
\frac{d}{dt}[\Delta q_S(x,t)]=\frac{\alpha_s(t)}{2\pi}
[\Delta P_{qq}\otimes\Delta q_S(x,t)+2N_f\Delta P_{qG}\otimes
\Delta g(x,t)]
\end{equation}
\begin{equation}\label{rA.3}
\frac{d}{dt}[\Delta g(x,t)]=\frac{\alpha_s(t)}{2\pi}
[\Delta P_{Gq}\otimes\Delta q_S(x,t)+2N_f\Delta P_{GG}\otimes
\Delta g(x,t)]
\end{equation}
where $t=\ln (Q^2/\Lambda^2)$ and
\begin{equation}\label{rA.4}
P(x)\otimes q(x,t)\equiv\int\limits_0^1\frac{dz}{z}P(z)q(
\frac{x}{z},t)
\end{equation}
Functions $C(x,\alpha_s)$ and $P_{ij}(x,\alpha_s)$ are calculated in
leading LO approximation or with the next correction to LO {\it i.e.}
in NLO approximation with respect to coupling constant $\alpha_s$:
\begin{equation}\label{rA.5}
LO:~~~P_{ij}(x,\alpha_s)=\alpha_sP_{ij}^{(0)}(x)
\end{equation}
\begin{equation}\label{rA.6}
NLO:~~~P_{ij}(x,\alpha_s)=\alpha_sP_{ij}^{(0)}(x)+
\alpha_s^2P_{ij}^{(1)}(x)
\end{equation}
Functions $P_{ij}(x)$ for the unpolarised case are different from
those $\Delta P_{ij}(x)$ for the polarised one \cite{b10}. In LO they
have a form:
\begin{equation}\label{rA.7}
P_{qq}^{(0)}(x)=\frac43\frac{1+x^2}{(1-x)_+}+2\delta (1-x)
\end{equation}
\begin{equation}\label{rA.8}
P_{qG}^{(0)}(x)=\frac12(x^2+(1-x)^2)
\end{equation}
\begin{equation}\label{rA.9}
P_{Gq}^{(0)}(x)=\frac43\frac{1+(1-x)^2}{x}
\end{equation}
\begin{equation}\label{rA.10}
P_{GG}^{(0)}(x)=3[\frac{x}{(1-x)_+}+\frac{1-x}{x}+x(1-x)
+\frac34\delta (1-x)]
\end{equation}
and
\begin{equation}\label{rA.11}
\Delta P_{qq}^{(0)}(x)=\frac43\frac{1+x^2}{(1-x)_+}
\end{equation}
\begin{equation}\label{rA.12}
\Delta P_{qG}^{(0)}(x)=\frac12(2x-1)
\end{equation}
\begin{equation}\label{rA.13}
\Delta P_{Gq}^{(0)}(x)=\frac43(2-x)
\end{equation}
\begin{equation}\label{rA.14}
\Delta P_{GG}^{(0)}(x)=3[\frac{1+x^4}{(1-x)_+}+(3-3x+x^2+x^3)
-\frac{7}{12}\delta (1-x)]
\end{equation}
where $(1-x)_+$ is defined as:
\begin{equation}\label{rA.15}
\int\limits_0^1\frac{f(x)dx}{(1-x)_+}\equiv
\int\limits_0^1\frac{f(x)-f(1)}{(1-x)}dx
\end{equation}

\section{Dynamical input parametrisations of the nonsinglet
polarised structure function $g_1^{NS}$}

In our calculations we adopt GRSV (Gl\"uck, Reya, Stratmann, Vogelsang)
\cite{b22} and GS (Gehrmann, Stirling) \cite{b23} parametrisations of
valence quarks $\Delta q_{val}$. We assume SU(3) flavour symmetric
scenario, where
\begin{equation}\label{rB.1}
\Delta\bar{u}=\Delta\bar{d}
\end{equation}
This assumption leads to formula (\ref{r4.4}):
\begin{equation}\label{rB.2}
g_1^{NS}\equiv g_1^p-g_1^n=\frac16(\Delta u-\Delta d)=
\frac16(\Delta u_{val}-\Delta d_{val})
\end{equation}
Input parametrisation of $\Delta u_{val}$ and $\Delta d_{val}$ have a
general form:
\begin{equation}\label{rB.3}
GRSV:~~~\Delta q_{val}=Nx^{a_2}x^{a_1-1}(1+Ax^b+Bx+Cx^{3/2})(1-x)^D
\end{equation}
\begin{equation}\label{rB.4}
GS:~~~\Delta q_{val}=N'x^{a'-1}(1+\gamma x+\rho\sqrt{x})(1-x)^{D'}
\end{equation}
where $N$, $N'$ are normalisation factors, implied by Bjorken and
Ellis-Jaffe sum rules (\ref{r2.11})-(\ref{r2.14}). These sum rules for
input scale $Q_0^2$ can be read as
\begin{equation}\label{rB.5}
a_3=\int\limits_0^1(\Delta u_{val}-\Delta d_{val})dx=1.257
\end{equation}
\begin{equation}\label{rB.6}
a_8=\int\limits_0^1(\Delta u_{val}+\Delta d_{val})dx=0.579
\end{equation}
(\ref{rB.5}) and (\ref{rB.6}) give immediately
\begin{equation}\label{rB.7}
\int\limits_0^1\Delta u_{val}dx=0.918
\end{equation}
\begin{equation}\label{rB.8}
\int\limits_0^1\Delta d_{val}dx=-0.339
\end{equation}
what allows to find $N$, $N'$ factors. The full set of input
parameters for GRSV and GS distributions is as follows:\hfill
\newline
{\bf GRSV:}\hfill
\newline
$Q_0^2=1~{\rm GeV}^2$, $\Lambda_{QCD}=232$ MeV
\begin{eqnarray}
{\rm for}~ \Delta u_{val}:&
N=1.964,~ a_1=0.573,~ a_2=0.16,~ b=0.175,\nonumber\\
&A=-0.583,~ B=1.723,~ C=3.436,~ D=3.486\nonumber\\
{\rm for}~ \Delta d_{val}:&
N=-0.162,~ a_1=0.376,~ a_2=0,~ b=0.529,\nonumber\\
&A=1.195,~ B=6.164,~ C=2.726,~ D=4.215\nonumber
\end{eqnarray}
{\bf GS:}\hfill\newline
$Q_0^2=4~{\rm GeV}^2$, $\Lambda_{QCD}=200$ MeV
\begin{eqnarray}
{\rm for}~ \Delta u_{val}:&
N'=1.741,~ a'=0.578,~ \gamma =9.38,~ \rho =-4.26,~ D'=3.73\nonumber\\
{\rm for}~ \Delta d_{val}:&
N'=-1.176,~  a'=0.666,~ \gamma =10.46,~ \rho =-5.10,~ D'=4.73\nonumber
\end{eqnarray}
In both GRSV and GS inputs we employ the LO fits.
Thus the input parametrisations have final forms:\\
\\
{\bf GRSV:}
\begin{equation}\label{rB.9}
\Delta u_{val}=1.964x^{-0.267}(1-0.583x^{0.175}+1.723x+3.436x^{3/2})
(1-x)^{3.486}
\end{equation}
\begin{equation}\label{rB.10}
\Delta d_{val}=-0.162x^{-0.624}(1+1.195x^{0.529}+6.164x+2.726x^{3/2})
(1-x)^{4.215}
\end{equation}
{\bf GS:}
\begin{equation}\label{rB.11}
\Delta u_{val}=1.741x^{-0.422}(1+9.38x-4.26\sqrt{x})(1-x)^{3.73}
\end{equation}
\begin{equation}\label{rB.12}
\Delta d_{val}=-1.176x^{-0.334}(1+10.46x-5.10\sqrt{x})(1-x)^{4.73}
\end{equation}


\begin{thebibliography}{99}

\bibitem{b1} EMC Collaboration: J.Ashman et al., {\it Phys. Lett.}
{\bf B206}, 364 (1988); {\it Nucl. Phys.} {\bf B328}, 1
     (1989).
\bibitem{b2} P.D.B.Collins, {\it An Introduction to Regge Theory and
High Energy Physics}, Cambridge University Press, Cambridge 1977.
\bibitem{b3} G.Altarelli, G.Parisi, {\it Nucl. Phys.} {\bf B126}, 298
(1977); E.G.Floratos, D.A.Ross, C.T.Sachrajda, {\it Nucl. Phys.}
{\bf B129}, 66 (1977); {\bf E139}, 545 (1978); {\it Nucl. Phys.}
{\bf  B152}, 493 (1979); G.Curci, W.Furmanski, R.Petronzio,
{\it Nucl. Phys.} {\bf B175}, 27 (1980); W.Furmanski, R.Petronzio,
{\it Phys. Lett.} {\bf 97B}, 437 (1980); M.A.Ahmed, G.G.Ross,
{\it Nucl. Phys.} {\bf B111}, 441 (1976).
\bibitem{b4} E.A.Kuraev, L.N.Lipatov, V.Fadin, {\it Zh. Eksp. Teor.
Fiz.} {\bf 72}, 373 (1977) {\it Sov. Phys. JETP} {\bf 45}, 199 (1977));
Ya.Ya.Balitzkij, L.N.Lipatov, {\it Yad. Fiz.} {\bf 28}, 1597 (1978)
{\it Sov. J. Nucl. Phys.} {\bf 28}, 822 (1978)); L.N.Lipatov, in
{\it Perturbative QCD}, edited by A.H.Mueller, World Scientific,
Singapore 1989, p.441; J.B.Bronzan, R.L.Sugar, {\it Phys. Rev.}
{\bf D17}, 585 (1978); T.Jaroszewicz, {\it Acta Phys. Pol.} {\bf B11}, 965 (1980).
\bibitem{b5} R.D.Ball, S.Forte, {\it Phys. Lett.} {\bf B335}, 77
(1994); {\bf B336}, 77 (1994); {\it Acta Phys. Pol.} {\bf B26}, 2097
(1995); R.D.Ball, A.DeRoeck, in the {\it Proceedings of the
International Workshop on Deep Inelastic Scattering and Related Phenomena
(DIS96)}, hep-ph/9609309 + refs.therein.
\bibitem{b6} G.P.Ramsey, {\it Prog. Part. Nucl. Phys.} {\bf39}, 599
(1997).
\bibitem{b7} J.D.Bjorken, {\it Phys. Rev.} {\bf 148}, 1467 (1966);
{\it Phys. Rev.} {\bf D1}, 1376 (1970); J.Ellis, R.L.Jaffe,
{\it Phys. Rev.} {\bf D9}, 1444 (1974); {\it Phys. Rev.} {\bf D10},
1669 (1974); G.Altarelli et al., {\it Acta Phys. Pol.} {\bf B29},
1145 (1998).
\bibitem{b8} A.De Roeck et al., hep-ph/9801300; A.De Roeck,
{\it Acta Phys. Pol.} {\bf B29}, 1343 (1998).
\bibitem{b9} V.N.Gribov, L.N.Lipatov, {\it Sov. J. Nucl. Phys.}
{\bf 15}, 438 and 675 (1972); Yu.L.Dokshitzer, {\it Sov. Phys. JETP}
{\bf 46}, 641 (1977).
\bibitem{b10} R.Mertig, W.L. van Neerven, {\it Z. Phys.} {\bf C70},
637 (1996); W.Vogelsang, {\it Phys. Rev.} {\bf D54},
     2023 (1996); {\it Nucl. Phys.} {\bf B475}, 47 (1996).
\bibitem{b11} Gaby R\"adel, {\it Acta Phys. Pol.} {\bf B29}, 1295 (1998);
M.Ruh, {\it Acta Phys. Pol.} {\bf B29}, 1275 (1998) + refs.therein.
\bibitem{b12} T.J.Ketel, {\it Acta Phys. Pol.} {\bf B29}, 1265 (1998)
+ refs.therein.
\bibitem{b13} B.L.Ioffe, hep-ph/9408291.
\bibitem{b14} SMC, D.Adams et al., {\it Phys. Lett.} {\bf B329},
399 (1994); {\bf B336}, 125 (1994); {\it Phys. Rev.} {\bf D56},
5330 (1997); B.Adeva et al., {\it Phys. Lett.} {\bf B412}, 414 (1997);
SMC, B.Adeva et al., {\it Phys. Lett.} {\bf B302}, 533 (1993);
D.Adams et al., {\it Phys. Lett.} {\bf B357}, 248 (1995); {\bf B396},
338 (1997).
\bibitem{b15} J.P.Nassalski, {\it Acta Phys. Pol.} {\bf B29},
1315 (1998).
\bibitem{b16} B.Lampe, E.Reya, hep-ph/9810270.
\bibitem{b17} Wu-Ki Tung, FERMILAB-PUB/88-135-T.
\bibitem{b18} A.De Rujula et al., {\it Phys. Rev.} {\bf 10}, 1649 (1974);
\bibitem{b19} NMC Collaboration, M.Arneodo et al., {\it Phys. Rev.}
{\bf D50}, R1 (1994); W.G. Seligman et al., {\it Phys. Rev. Lett.}
{\bf 79}, 1213 (1997) + refs therein.
\bibitem{b20} J.Kwieci\'nski, {\it Acta Phys. Pol.} {\bf B29}, 1201
(1998).
\bibitem{b21} J.Soffer, {\it Acta Phys. Pol.} {\bf B29}, 1303 (1998),
L.C.Bland, hep-ex/9907058.
\bibitem{b22} M.Gl\"uck, E.Reya, M.Stratmann, W.Vogelsang,
{\it Phys. Rev.} {\bf D53}, 4775 (1996); M.Gl\"uck, E.Reya, A.Vogt,
{\it Z. Phys.} {\bf C67}, 433 (1995).
\bibitem{b23} T.Gehrmann, W.J.Stirling, {\it Phys. Rev.} {\bf D53},
6100 (1996).
\bibitem{b24} J.Bartels, B.I.Ermolaev, M.G.Ryskin, {\it Z. Phys.}
{\bf C70}, 273 (1996); J.Bartels, B.I.Ermolaev, M.G.Ryskin,
{\it Z. Phys.} {\bf C72}, 627 (1996); J.Kwieci\'nski,
{\it Acta Phys. Pol.} {\bf B27}, 893 (1996).
\bibitem{b25} B.Bade\l{}ek, J.Kwieci\'nski, {\it Phys. Lett.}
{\bf B418}, 229 (1998).
\bibitem{b26} M.Ruh, {\it Acta Phys. Pol.} {\bf B29}, 1275 (1998).

\end{thebibliography}
\end{document}